\documentclass[conference]{IEEEtran}
\usepackage[utf8]{inputenc}

\usepackage[annataritalic]{tengwarscript}

\usepackage{amsmath,amssymb,amsfonts}
\usepackage{algorithm}
\usepackage{algcompatible}
\usepackage{algpseudocode}
\usepackage{dblfloatfix}
\usepackage{hhline}
\usepackage{amsmath}
\usepackage{array, makecell, cellspace}
\usepackage{hyperref}

\usepackage[normalem]{ulem}
\usepackage{multirow}
\usepackage{multicol}
\usepackage{array}
\usepackage{xcolor}

\usepackage{graphicx}

\usepackage{titlesec}

\usepackage{booktabs}
\usepackage{soul}

\makeatletter
\let\@ORGmakecaption\@makecaption
\long\def\@makecaption#1#2{\@ORGmakecaption{#1}{#2}\vskip\belowcaptionskip\relax}
\makeatother
\usepackage{xspace}

\usepackage{pifont}

\usepackage{listings}
\usepackage{xcolor} 
\usepackage{soul} 
\usepackage{listings} 
\usepackage{float}

\definecolor{lightred}{rgb}{1, 0.8, 0.8}

\lstdefinestyle{mystyle}{
    language=C,
    breaklines=true,
    numbers=left, 
    numberstyle=\scriptsize,
    stepnumber=1,
    numbersep=12pt, 
    frame=single,
    basicstyle=\small\ttfamily,
    xleftmargin=20pt, 
    showstringspaces=false,
    backgroundcolor=\color{white}, 
    literate={^}{{\^{}}}1,  
}

\newcommand{\hlcode}[1]{\colorbox{lightred}{\strut #1}}

\usepackage{lipsum}

\definecolor{mGreen}{rgb}{0,0.6,0}
\definecolor{mGray}{rgb}{0.5,0.5,0.5}
\definecolor{mPurple}{rgb}{0.58,0,0.82}
\definecolor{backgroundColour}{rgb}{0.95,0.95,0.92}

\lstdefinestyle{CStyle}{
    backgroundcolor=\color{backgroundColour},   
    commentstyle=\color{mGreen},
    keywordstyle=\color{magenta},
    numberstyle=\tiny\color{mGray},
    stringstyle=\color{mPurple},
    basicstyle=\footnotesize,
    breakatwhitespace=false,         
    breaklines=true,                 
    captionpos=b,                    
    keepspaces=true,                 
    numbers=left,                    
    numbersep=5pt,                  
    showspaces=false,                
    showstringspaces=false,
    showtabs=false,                  
    tabsize=2,
    language=C
}

\begin{document}

\setstcolor{red}

\title{SHIFT SNARE: Uncovering Secret Keys in FALCON via Single-Trace Analysis\\}

\author{

\IEEEauthorblockN{
Jinyi Qiu$^{\dag}$, Aydin~Aysu$^{\dag}$}
\IEEEauthorblockA{$^{\dag}$Department of Electrical and Computer Engineering, North Carolina State University, Raleigh, NC, USA 
}
}

\maketitle            

\vspace{+5em}
\thispagestyle{plain}
\pagestyle{plain}

\begin{abstract}

This paper presents a novel \emph{single-trace} side-channel attack on FALCON, a lattice-based post-quantum digital signature protocol recently approved for standardization by NIST. 
We target the discrete Gaussian sampling operation within FALCON's key generation scheme and demonstrate that a single power trace is sufficient to mount a successful attack.
Notably, negating the results of a 63-bit right-shift operations on 64-bit secret values leaks critical information about the assignment of ‘-1’ versus ‘0’ to intermediate coefficients during sampling. 
These leaks enable full recovery of the secret key.

We demonstrate a ground-up approach to the attack on an ARM Cortex-M4 microcontroller executing both the reference and optimized implementations from FALCON’s NIST round 3 software package. 
We successfully recovered all of the secret polynomials in FALCON.
We further quantify the attacker’s success rate using a univariate Gaussian template model, providing generalizable guarantees.
Statistical analysis with over 500,000 tests reveals a per-coefficient success rate of 99.9999999478\% and a full-key recovery rate of 99.99994654\% for FALCON-512. 
We verify that this vulnerability is present in all implementations included in FALCON’s NIST submission package.
This highlights the vulnerability of current software implementations to single-trace attacks and underscores the urgent need for single-trace-resilient software in embedded systems.

\begin{IEEEkeywords}
Side-channel attacks, Post-quantum cryptography, NTRU, FALCON, Key generation, Lattice-based cryptography, Digital signature schemes
\end{IEEEkeywords}

\end{abstract}

\section{Introduction} 
Widely adopted encryption schemes such as RSA~\cite{RSA} and elliptic curve cryptosystems~\cite{ECC} rely on hard mathematical problems,
including integer factorization~\cite{montgomery1994survey} and the discrete logarithm problem~\cite{diffie_new_1976}, 
which are traditionally regarded as computationally intractable for classical computers. 
However, quantum algorithms offer exponential speedups for solving these problems~\cite{Shor1999}. 
As a result, advances in quantum computing pose a significant threat to conventional encryption schemes and underscore the critical need to design cryptographic systems that can resist quantum attacks.

To address this issue, the National Institute of Standards and Technology (NIST) initiated a standardization process for post-quantum cryptographic schemes, also referred to as quantum-resistant algorithms, designed to withstand quantum cryptanalysis~\cite{NIST_call}. 
As of now, this process has selected three digital signature schemes for standardization: CRYSTALS-Dilithium~\cite{lyubashevsky2020crystals}, SPHINCS+~\cite{bernstein2019sphincs+}, and FALCON~\cite{prest2020falcon}.
NIST selected FALCON in part because of its small signature size, making it particularly suitable for embedded and bandwidth-constrained systems.

While the algorithms chosen are expected to be mathematically robust, their practical implementations may remain vulnerable to side-channel attacks. 
These attacks exploit physical characteristics of implementations, such as execution time, power consumption, and electromagnetic emissions, to extract secret information~\cite{dpa}. An attacker can carry out such attacks using only a few side-channel measurements from the physical device~\cite{ravi_generic, malik2024enabling, kurian2025tpuxtract}. 
The most severe form of these attacks, known as\textit{ single-trace} attacks or simple power analysis, enables adversaries to recover secret data from just a single execution of the program. 
Single-trace attacks are particularly dangerous because they can bypass commonly deployed defenses such as masking~\cite{ngo2021side}. 
Furthermore, they can target sensitive subroutines such as key generation, which produces a new secret on each invocation.
FALCON’s suitability for embedded deployment makes it a prime target for side-channel exploitation. 
Given the imminent real-world deployment of NIST’s post-quantum cryptographic standards \cite{sedghighadikolaei2023comprehensive, liu2024road, li2023trends}, there is a critical need to uncover such vulnerabilities and guide the development of effective countermeasures.

Previous work on \textit{single-trace} side-channel analysis of lattice-based cryptosystems has identified several vulnerable components. 
Examples include the number-theoretic transform (NTT)~\cite{first_NTT_attack, second_NTT_attack}, polynomial multiplication~\cite{aysu18horizontal, frodo_single_poly}, message encoding/decoding~\cite{message_recovery, rabas2024single}, cumulative distribution table (CDT) sampling~\cite{single_trace_18, choi2024single}, and the Fujisaki–Okamoto transform \cite{jendral2024breaking}.
Although FALCON incorporates some of these components, it also introduces unique elements such as fast Fourier sampling and floating-point arithmetic. 
Existing attacks cannot be directly applied to these FALCON-specific operations, highlighting the need for further investigation into vulnerabilities specific to FALCON.

\begin{figure}[t]
    \centering
    \includegraphics[scale=0.33]{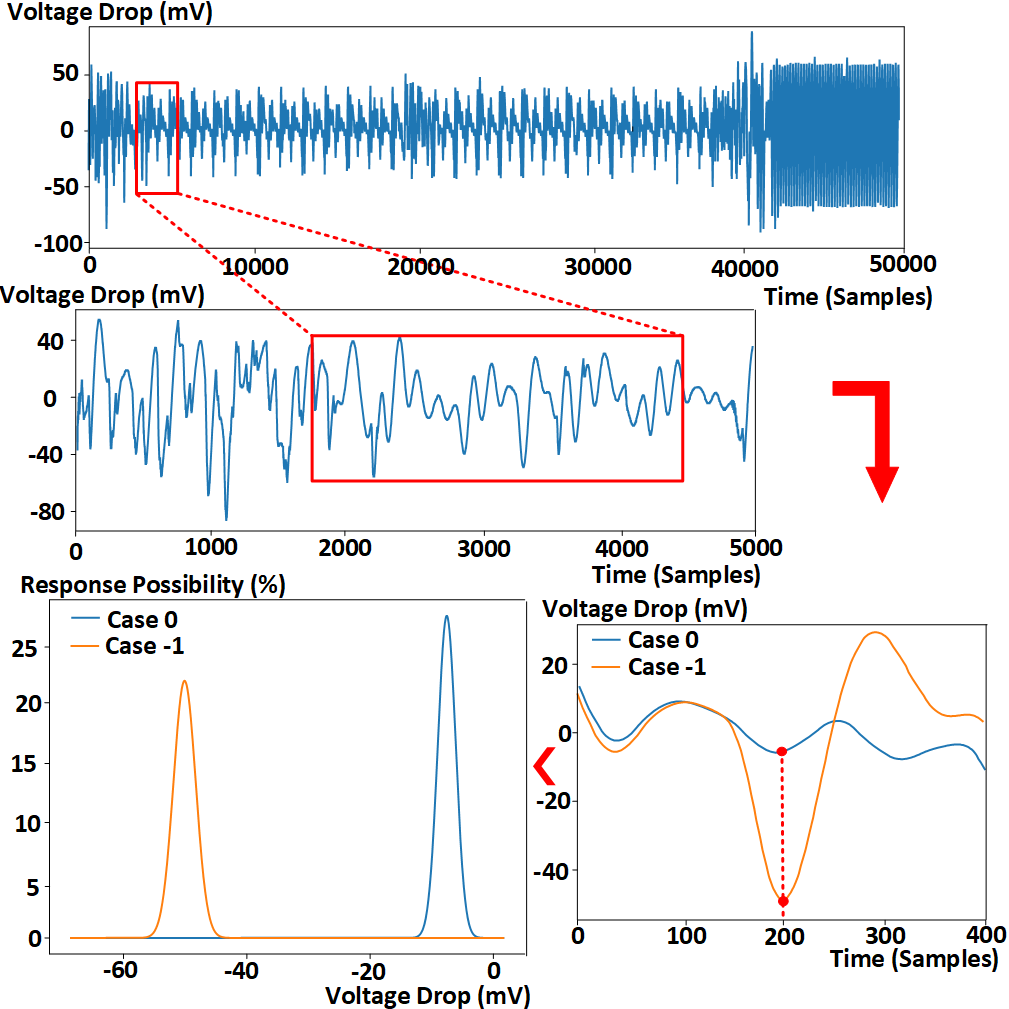}
    \vspace{-1.75em}
    \caption{\small Visual Demonstration of the Vulnerability: The top figure illustrates the power consumption profile of the device during the key generation process. The middle figure provides a zoomed-in view of the power consumption during a specific vulnerable segment in the code. The bottom right figure offers a further magnified view and the average power consumption for two distinct cases of secret value assignment (`0' versus `-1'). The bottom left figure presents the results of a univariate Gaussian model distinguishing these two classes over 500,000 trials. The analysis reveals that different assignments of secret intermediate values cause significant variations in power consumption. This demonstrates the practicality of mounting an attack with a substantial success rate.}
    \vspace{.5em}
    \label{fig:G1} 
\end{figure}

In this paper, we present a \textit{novel} single-trace side-channel vulnerability in FALCON that is distinct from previously reported attacks~\cite{choi2024single, guerreau2022hidden, zhang2023improved, lin2025thorough, marzougui2022machine, schonauer2024physical}. 
\autoref{fig:G1} provides a visual summary of the vulnerability and emphasizes its practical relevance. 
The top panel displays a power trace recorded during the discrete Gaussian sampling phase of FALCON key generation. 
The middle panel highlights the region of interest where data-dependent leakage is apparent. 
The bottom right panel presents a magnified view, illustrating the average power consumption corresponding to two secret value assignments: `0' and `-1'. 
The bottom left panel shows the output of a univariate Gaussian model evaluated at the point of maximum leakage, clearly demonstrating a separation between the two classes.
This separation confirms the presence of an exploitable vulnerability in FALCON’s implementation.

The contributions of this paper are as follows:
\begin{itemize}
    \item We demonstrate a \textit{new} side-channel vulnerability in FALCON’s key generation process that enables full recovery of the secret key from a single power measurement.
    
    \item We present a custom method developed from first principles to recover the entire secret key. A theoretical model is constructed to provide a generalizable guarantee on the success rate.
    
    \item We apply the attack to an off-the-shelf device featuring an ARM Cortex-M4 microcontroller—a widely used platform for side-channel evaluation. The microcontroller executes the reference implementation from FALCON’s NIST submission package. Our practical experiments achieve a 100\% success rate, while the theoretical model predicts a per-variable success rate of 99.9999999478\% and a full key recovery rate higher than 99.99989309\%. The attack is effective across all implementations within the NIST submission package, as the same leakage is consistently observed in both the reference and optimized versions.

     \item To support reproducibility and future research, we will publicly release our power traces and attack code upon acceptance of this paper.
\end{itemize}

\vspace{-0.8em}
\section{background}
\label{sec:pre}
This section provides an overview of FALCON and explains the role of its secret polynomials. 
We also highlight the differences between our attack and previous side-channel attacks on FALCON. 
Finally, we present our threat model.

\subsection{The Generation of FALCON's Secret Polynomial}
FALCON (Fast Fourier Lattice-based Compact Signature over NTRU) is a digital signature scheme designed for the post-quantum era. 
It provides security against quantum adversaries, as even quantum computers would require infeasible computational resources to break the mathematical trapdoors underlying the scheme. 
This resilience is derived from the Ring Learning with Errors (R-LWE) problem, which FALCON incorporates within the NTRU lattice framework.
In comparison to other post-quantum algorithms, FALCON features relatively small key and signature sizes, enhancing its efficiency and making it well-suited for deployment in low-power devices and bandwidth-constrained environments.

FALCON comprises three main stages: key generation, signature generation, and signature verification. 
This work focuses on the key generation step, specifically the computation of the secret polynomials $f$ and $g$.

\begin{algorithm}
  \caption{FALCON Key Generation}
  \label{alg:FYShuffle}		
    \begin{algorithmic}[1]
      \Require A monic polynomial $\phi \in \mathbb{Z}[x]$, a modulus $q$ 
      \Ensure  A secret key sk, a public key pk
  \end{algorithmic}%
  \protect
  \begin{algorithmic}[1]
        \State $f, g, F, G \gets \text{NTRUGen}(\phi, q)$ \Comment{Solving the NTRU equation}
        \State $\mathbf{B} \gets \begin{bmatrix} g & -f \\ G & -F \end{bmatrix}$
        \State $\hat{\mathbf{B}} \gets \text{FFT}(\mathbf{B})$ \Comment{Compute the FFT for each of the 4 components $\{g, -f, G, -F\}$}
        \State $\mathbf{G} \gets \hat{\mathbf{B}} \times \hat{\mathbf{B}}^\star$
        \State $\mathbf{T} \gets \text{ffLDL}^\star(\mathbf{G})$ \Comment{Computing the LDL$^\star$ tree}
       \State \textbf{for each} leaf of $\mathbf{T}$ \textbf{do} \Comment{Loop over the leaves of $\mathbf{T}$}
        \State \hskip1em leaf.value $\gets \sigma / \sqrt{\text{leaf.value}}$ \Comment{Normalization step}
        \State sk $\gets (\hat{\mathbf{B}}, \mathbf{T})$
        \State $h \gets g f^{-1} \mod q$
        \State pk $\gets h$
        \State \textbf{return} sk, pk
    \end{algorithmic}
\end{algorithm}

\autoref{alg:FYShuffle} illustrates the steps involved in FALCON’s key generation process. 
It begins with a predefined parameter $n$ (set to either 512 or 1024) and a modulus $q$ (set to 12289 in the NIST submission package), which define the size of the ring.
Initially, the base NTRU lattice components, $f$ and $g$, are generated. 
These components are then used to deterministically derive both the public and secret keys—without relying on randomness or additional secret variables. 
This deterministic derivation implies that \textbf{the base NTRU lattice components, $g$ and $f$, are critically important, since both keys are fixed once $f$ and $g$ are generated.} 
Although the public key $h$ is derived from these components, the process is non-invertible, meaning an adversary cannot reconstruct the base lattice components $f$ and $g$ from the public key $h$. 
The remainder of this subsection provides a brief explanation of how FALCON generates the public key $pk$ and secret key $sk$, to contextualize the attack described in \autoref{sec:LeakyOps}.

\begin{algorithm}[!tb]
  \caption{NTRUGen$(\phi, q)$}
  \label{alg:NTRUGen}
\begin{algorithmic}[1]
\Require A monic polynomial $\phi \in \mathbb{Z}[x]$ of degree $n$, a modulus $q$
\Ensure Polynomials $f, g, F, G$
\State $\sigma_{(f,g)} \gets 1.17 \sqrt{q}/2n$ \Comment{$\sigma_{(f,g)}$ is chosen so that $\mathbb{E}[\| (f, g) \|] = 1.17\sqrt{q}$}
\For{$i$  from  $0$ to $n-1$}
    \State $f_i \gets D_{\mathbb{Z}, \sigma_{(f,g)}}$
    \State $g_i \gets D_{\mathbb{Z}, \sigma_{(f,g)}, 0}$
\EndFor
\State $f \gets \sum_i f_i x^i$ \Comment{$f \in \mathbb{Z}[x]/(\phi)$}
\State $g \gets \sum_i g_i x^i$ \Comment{$g \in \mathbb{Z}[x]/(\phi)$}

...

\State $F, G \gets \text{NTRUSolve}_n, q (f, g)$ \Comment{Computing $F, G$ such that $f G - g F = q \mod \phi$}
\If{$(F, G) = 1$}
    \State \textbf{restart}
\EndIf
\State \textbf{return} $f, g, F, G$
\end{algorithmic}
\end{algorithm}

\autoref{alg:NTRUGen}, $NTRUGen()$, is the first step in generating the base NTRU lattice components $f$ and $g$ based on a Gaussian distribution. 
Lines 2 through 7 of the algorithm define $f$ and $g$ as polynomials of degree $n$, following the structure described in the equation below:
\vspace{-.25em}
\begin{equation} \label{eq:f_poly}
f(x) = f_0 + f_1 x + f_2 x^2 + f_3 x^3 + \dots + f_{n-1} x^{n-1}
\end{equation}
\vspace{-.25em}
\begin{equation} \label{eq:g_poly}
g(x) = g_0 + g_1 x + g_2 x^2 + g_3 x^3 + \dots + g_{n-1} x^{n-1}
\end{equation}

In the formula above, the coefficients of $f$ and $g$ are sampled discretely from a Gaussian distribution. 
The mean of this distribution is zero, and the standard deviation is determined by the degree $n$ and the parameter $q$, as specified in line 1 of \autoref{alg:NTRUGen}. 
For FALCON-512, $n$ is set to 512, whereas for FALCON-1024, $n$ is set to 1024. In both parameter sets, the modulus $q$ is fixed at 12289.

In line 8 of \autoref{alg:NTRUGen}, after generating the lattice base components, the key generation process solves the NTRU equation (\autoref{eq:NTRU_eq}) using $f$ and $g$ to derive the secret key and the public key. 
\begin{equation} \label{eq:NTRU_eq}
f G - g F = q \mod \phi
\end{equation}
To obtain $F$ and $G$, which are essential for generating the public and secret keys, the algorithm recursively reduces the polynomials $f$ and $g$ into two polynomials whose degrees are half those of their predecessors. 
When the polynomial degree is reduced to 1, the base elements of $F$ and $G$ are obtained by finding the greatest common divisor (GCD) of $f$ and $g$. 
Once the base elements of $F$ and $G$ are established, they are recursively combined, doubling the degree of the resulting polynomial at each step.
This process continues until the polynomials $F$ and $G$ are reconstructed at degree $n$.

The remainder of the key generation process involves computing the fast Fourier transform (FFT), constructing the Gram matrix, and generating the FALCON tree based on this matrix. 
Notably, this portion of the process is deterministic and does not introduce additional randomness, provided that the NTRU lattice components $f$, $F$, $g$, and $G$ are known. 
\subsection{Comparison to Previous Attacks on FALCON}

Previous studies investigating side-channel attacks on FALCON include the work of Karabulut et al.~\cite{karabulut2021falcon}, who presented the first multi-trace side-channel attack on the scheme, and McCarthy et al.~\cite{mccarthy2019bearz}, who introduced the first fault injection attack. 
FALCON’s vulnerability to single-trace attacks has also been examined at the algorithmic level; for example, Guerreau et al. \cite{guerreau2022hidden} analyzed leakage in the base sampling procedure, while Zhang et al. \cite{zhang2023improved} proposed improvements to this approach. 
More recently, Lin et al.~\cite{lin2025thorough} identified side-channel leakage originating from the half-Gaussian sampler used during FALCON’s signature generation.
The vulnerability we identify is at a different subroutine and thus is orthogonal to these single-trace FALCON attacks.

We discovered a novel side-channel vulnerability in a previously unexplored stage of the FALCON signature scheme---key generation. 
The root cause of this vulnerability arises from the leaky negation operations, initially identified by Karabulut et al.~\cite{karabulut2021single} and later leveraged by Guerreau et al.~\cite{guerreau2024not}. 
However, our identification of how this vulnerability manifests in the targeted subroutine and our corresponding exploitation techniques for full secret-key recovery constitute the novel contributions of this work.

\subsection{Threat Model}
\label{sec:T_Model}
We adopt the well-established threat model for single-trace side-channel attacks \cite{lin2025thorough, karabulut2021single,guerreau2024not}, assuming an adversary who has physical access to the target device and can measure its power consumption during cryptographic operations. 
The adversary is assumed to possess knowledge of the executing software, to approximate the timing of specific computations, and to intercept communication channels to capture exchanged public messages. 
Our attack uses a profiling phase. 
During this phase, the attacker can supply random inputs and analyze the software’s power consumption behavior using known values. 
However, at runtime, the adversary is restricted to capturing a single power trace in an attempt to deduce the entire secret key. 

We conduct our experiments on the STM32F417 development board, which features an ARM Cortex-M4 processor, one of the most widely used platforms in side-channel analysis research \cite{  guerreau2022hidden, zhang2023improved, lin2025thorough, karabulut2021falcon, karabulut2021single, guerreau2024not}.
Although evaluating the vulnerability across multiple hardware platforms is a reasonable direction for future work, we focus exclusively on this target device, following the approach taken in prior studies \cite{kurian2025tpuxtract, lin2025thorough, karabulut2021single, guerreau2024not}.

The design and implementation of countermeasures are outside the scope of this paper, which focuses primarily on attack methodologies, following the approach taken in prior work \cite{karabulut2021falcon, karabulut2021single}. Other attacks, such as fault injection~\cite {malik2025craft}, are likewise considered out of scope.
\section{Underlying mechanism of the Attack}
\label{sec:LeakyOps}
In this section, we first demonstrate how negating the result of the target bit-shift operation leaks intermediate secret variables. 
We also provide proof-of-concept demonstrations that confirm this vulnerability. 
Subsequently, we explain how these leaked variables enable the full recovery of FALCON’s secret polynomial.

\setlength{\abovecaptionskip}{0pt} 
\setlength{\belowcaptionskip}{1pt} 
\begin{figure}[t]
    \centering
    \includegraphics[scale=0.28]{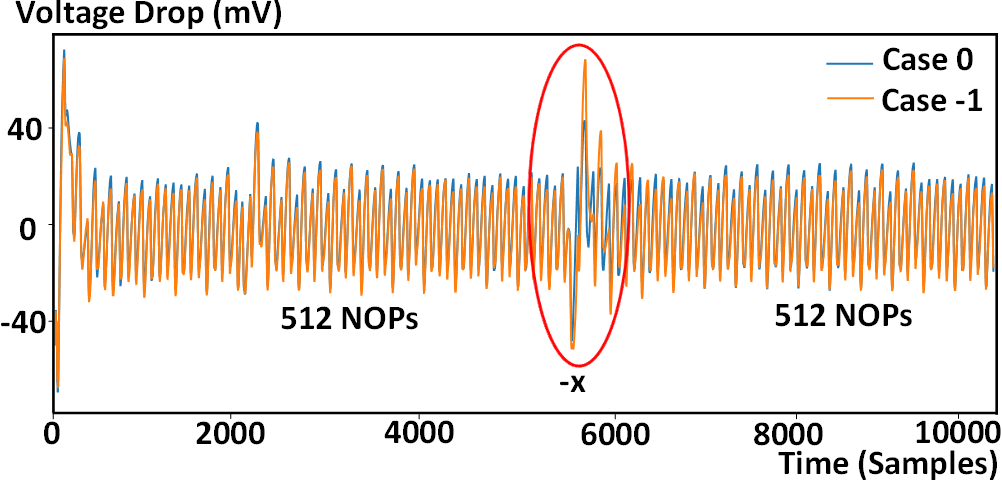}
    \caption{Power traces of the leaky operation under varying inputs are presented. The blue trace depicts the power consumption during the leaky operation when the secret is assigned the value `0', while the orange trace shows the power consumption when the secret is assigned `-1'. A distinct difference in power consumption between the two cases is observable.}
    \vspace{1em}
    \label{fig:preliminary} 
\end{figure}

\subsection{The Operations That Leak}
\label{sec:Ops}
This section describes how negating the `63-bit right-shift' operation can leak whether the result is `0' or `-1', and we present preliminary results to substantiate this claim.

For a 64-bit variable, the `63-bit right-shift' operation (expressed as `\( (x \gg 63) \)' in software) shifts the most significant bit (MSB) to the least significant bit (LSB) and clears all other bits. 
This operation produces only two possible outcomes: `0' or `1'. 
Negating the result yields either `0' or `-1', which are represented in 64-bit two’s complement as all 0s and all 1s, respectively.
When the processor writes the result, the `-1' case exhibits a Hamming weight (HW) of 64, while the `0' case has an HW of 0. 
\textbf{Consequently, the `-1' case results in higher power consumption compared to the `0' case.}
This leakage was first identified by Karabulut et al. ~\cite{karabulut2021single}. 
Guerreau et al. ~\cite{guerreau2024not} extended the analysis of this leakage.
\textbf{However, this is the first study to identify this vulnerability in FALCON's official software submission package.}

\autoref{fig:preliminary} presents experimental results validating the observed power difference caused by this operation.  
The experiment involves performing a negation operation on a 64-bit value using the assembly code shown in \autoref{alg:Asm}, aligning with the FALCON implementation. 
The \texttt{sbc.w} instruction is employed for two’s-complement negation (sign inversion) of the 64-bit value. 
The power consumption of the operation \( -(x) \) was measured, with \( x \) taking a value of either `1' or `0' (the result of \( x\gg63 \)). 
Assembly \texttt{NOP} instructions were inserted around the operation to isolate it. 
The resulting traces were overlaid for comparison. 
The blue trace corresponds to \( x \) is `0' and shows a peak voltage drop of 40 mV, whereas the orange trace, corresponding to \( x \) is `1', exhibits a drop of 70 mV. 
Furthermore, the power spike is more pronounced in the `-1’ case compared to the `0' case, demonstrating the vulnerability.
\begin{figure}[t]
\vspace{-1em} 
\centering
\begin{lstlisting}[style=mystyle, caption={Assembly instructions corresponding to \( -(x)\)}, label={alg:Asm}]
negs	r2, r2;
sbc.w	r3, r3, r3, lsl #1;
strd	r2, r3, [r7, #24];
\end{lstlisting}
\end{figure}

\begin{figure}[t]
\centering
\begin{lstlisting}[style=mystyle, caption={Gaussian sampling implementation from NIST submission package}, label={alg:gaussian_sampling}]
mkgauss(RNG_CONTEXT *rng, unsigned logn){
    ...
    for (u = 0; u < g; u ++) {
        ...
        r = get_rng_u64(rng);
        neg = (uint32_t)(r >> 63);
        r &= ~((uint64_t)1 << 63);
        f = (uint32_t)((r - gauss_1024_12289[0]) >> 63);
        v = 0;
        r = get_rng_u64(rng);
        r &= ~((uint64_t)1 << 63);
        for (k = 1; k < (sizeof gauss_1024_12289) 
            / (sizeof gauss_1024_12289[0]); k ++){
            uint32_t t;
            t = (uint32_t)((r - gauss_1024_12289[k]) >> 63) ^ 1;
            (*@\hlcode{ v |= k \& -(t \& (f \textasciicircum  1));}@*)  
            f |= t;}
         (*@\hlcode{v = (v \textasciicircum  -neg) + neg;}@*)
        val += *(int32_t *)&v;}
    return val;}
  \end{lstlisting}
  \vspace{1em}
\end{figure}
\vspace{-.5em}
\subsection{Only Two Variables Needed}
\label{sec:Getkey}
This subsection explains how an adversary can recover FALCON's base components $f$ and $g$ using only two intermediate variables within the discrete Gaussian sampling subroutine. 
Recall that the coefficients of $f$ and $g$ are generated using a discrete Gaussian sampling process. 
The reference C implementation of this process from the NIST submission package is outlined in \autoref{alg:gaussian_sampling}. 
This subroutine is called $n$ times to generate values for a degree-n polynomial forming the NTRU base components.
The parameter $n$ is the degree of the polynomial specified by the user. 
In the NIST submission package, $n$ is configured as 512 or 1024. 
The implementation consists of an outer and an inner loop. 
The number of outer loop executions depends on the value of the variable \texttt{logn}. 
The number of inner loop executions depends on both the dimensions and contents of the predefined matrix \texttt{Gauss\_1024\_12289}.
Static analysis reveals that the outer loop executes twice, and the inner loop executes 26 times per outer iteration.
We subsequently audited the source code to identify how this behavior leads to side-channel leakage.

Within \autoref{alg:gaussian_sampling}, line 20 returns \texttt{val}, which holds the generated secret coefficient. 
To extract the generated secret coefficient, we trace the changes to this variable back in this subroutine. 
Line 19 performs an update that, despite involving different access patterns, simplifies to \texttt{val =  val + v}. 
Since \texttt{val} is initialized to 0, the adversary only needs to infer the value of v for each outer loop to deduce the returned result. 
In Line 18, \texttt{v} is XORed with \texttt{-neg} and then added to \texttt{neg}, implying that the adversary also needs to know \texttt{neg} to determine \texttt{val}. 
Therefore, we identify two critical points in the subroutine that enable the recovery of $f$ and $g$. 
Specifically, line 16 reveals the value of \texttt{v}, and line 18 reveals \texttt{neg}, as highlighted in the listing. 

Line 16 is our first attack point because the value of \texttt{-(t \& (f \^{} 1))} can only evaluate to `0' or `-1'. 
The reason is that \texttt{t} and \texttt{f}—which are local variables distinct from the base component $f$—can only take `0' or `1' as a result of the `63-bit shift' operation on lines 8 and 15, which retains only the most significant bit (MSB).
The negation of \texttt{t \& (f \^{} 1)} thus simplifies to the negation of `0' or `1', respectively. 
The negation result, expressed in two's complement, will be all zeros (when the result is `0') or all ones (when the result is `-1'). 
\textbf{This will cause a significant difference in power consumption in the target device because the Hamming weight (HW) of these two results differs by 64.} 
Additionally, in line 16, $k$ represents the iteration index of the inner loop and ranges from 1 to 26. 
An adversary can count recurring power peaks to infer the value of \texttt{k}. 

Line 18 is our second attack point because \texttt{-neg} can only take `0' or `-1'. 
This is due to the `63-bit shift' operation on line 6 of the subroutine. 
The negation of \texttt{neg} introduces a vulnerability analogous to the one described at line 16, due to the stark difference in Hamming weight (HW). 
An adversary can infer the value of \texttt{neg} by exploiting this vulnerability.

Since both points of exploitation are located within a loop, each loop iteration depends on the value of \texttt{v} generated during previous iterations. 
Therefore, the attack must achieve a high success rate to recover the secret correctly.
Any incorrect inference of an intermediate value may result in a deviation from the correct output.
\begin{figure}[t]
    \centering
     \includegraphics[scale=0.28]{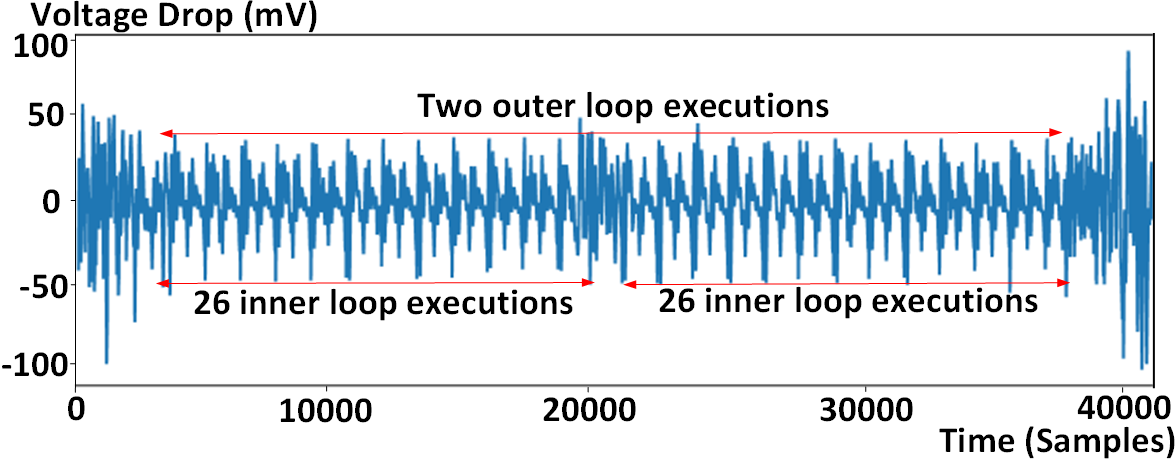}

    \caption{The full power trace of the subroutine running on the target device (an STM32F417 development board containing an ARM Cortex-M4 CPU) is shown. The two outer-loop executions and the 26 inner-loop executions are clearly observable.}
      \vspace{1.5em}
    \label{fig:Full} 
\end{figure}

\section{Exploiting the Found Vulnerability}
\label{sec:attack}
This section presents the proposed attack strategy to recover the secret polynomials in FALCON.
The power trace is analyzed to identify points of interest associated with information leakage. 
As noted previously, the attack begins with a profiling stage, where the adversary has physical access to the target device and knowledge of the software implementation.
In this stage, power measurements are collected under varying software inputs to construct a leakage profile to determine \textit{when} leakage occurs. 
However, after profiling is complete, the adversary can recover the secret polynomial from a single power measurement.

\subsection{Inspecting the Power Trace}
\autoref{fig:Full} shows the full power trace obtained when executing the discrete Gaussian sampling subroutine. 
Two outer loop executions and 26 inner loop executions per outer iteration are clearly distinguishable, reflecting the structure of the code in \autoref{alg:gaussian_sampling}. 
Each inner loop iteration produces a recurring pattern every 700 samples, while each outer loop spans 19000 samples.
Because the discrete Gaussian sampling subroutine is executed $n$ times during key generation, the resulting power traces exhibit a consistent and easily identifiable structure.

\subsection{Pinpointing the Point of Interest}
We use the following approach to apply correlation power analysis (CPA) to identify the point-of-interest (POI). 
The Pearson correlation coefficients are computed between the power measurements and a predetermined value set over time. 
First, we assign values to the variable \texttt{r} and adjust the values in the matrix \texttt{gauss\_1024\_12289} so that the value of  in the first and third inner loop executions can be manually controlled. 
We then collect 500,000 measurements, with \texttt{-(t \& (f \^{} 1))} set to `-1' in the first inner loop and `0' in the third inner loop for the first 250,000 measurements. 
For the second 250,000 measurements, \texttt{-(t \& (f \^{} 1))} is set to `0' in the first inner loop and `-1' in the third inner loop. 
The predetermined value set also contains 500,000 numbers. 
The first 250,000 values are set to 64, and the remaining 250,000 are set to 0, corresponding to the assigned values of \texttt{-(t \& (f \^{} 1))} in each case. 

When \texttt{-(t \& (f \^{} 1))} is assigned the value `-1', it will have a Hamming weight (HW) of 64. 
When \texttt{-(t \& (f \^{} 1))} is assigned the value `0', it will have an HW of 0. 
Due to the power characteristics described in ~\autoref{sec:LeakyOps}, this difference in Hamming weight is reflected in the device’s power consumption. 
This leads to a strong correlation between the device's power consumption and the predetermined value set at the timestamp when \texttt{-(t \& (f \^{} 1))} occurs in the first and third inner loop executions. 
\textbf{Consequently, the Pearson correlation coefficient peaks at these two timestamps.} 

\begin{figure}[t]
    \centering
    \includegraphics[scale=0.265]{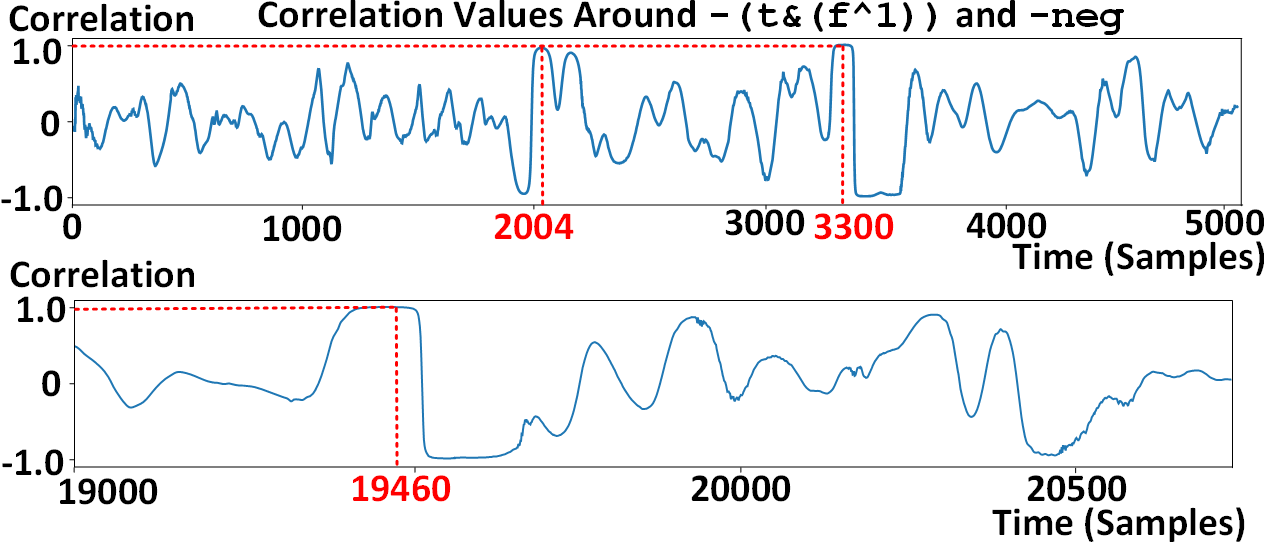}
    ~\vspace{-1.5em}
    \caption{Correlation power analysis (CPA) results: For the first attack point, the leaky operation occurs around timestamps 2004 and 3300, corresponding to correlation values of 0.996 and 0.977. For the second point, the leaky operation occurs around timestamp 19460, corresponding to a correlation value of 0.992.}
      \vspace{1em}
    \label{fig:Corrs} 
\end{figure}

Similarly, for the second leaky operation, we controlled the conditions under which \texttt{-neg} is computed and applied Pearson correlation between the power traces and the predetermined value set. 
We focused on the power trace segment between the end of the last inner loop and the start of the second outer loop execution.
The Pearson correlation coefficient peaks at the timestamp when \texttt{-neg} is computed. 

\autoref{fig:Corrs} (top) illustrates the Pearson correlation observed over time for the first leaky operation, \texttt{-(t \& (f \^{} 1))}, and Figure 4 (bottom) shows the result for the second leaky operation, \texttt{-neg}. 
We found that the timestamps with the highest correlation occur around samples 2004 and 3300, corresponding to when \texttt{-(t \& (f \^{} 1))} was computed in the first and third inner loop executions.  
For \texttt{-neg}, we found that the highest correlation timestamps occur around sample 19460. 
Although a few other timestamps also exhibit elevated correlation,~\autoref{sec:res} demonstrates that the chosen attack points are sufficient to extract the full key with high accuracy.

\begin{figure}[t]
	\centerline{\includegraphics[width=0.47\textwidth]{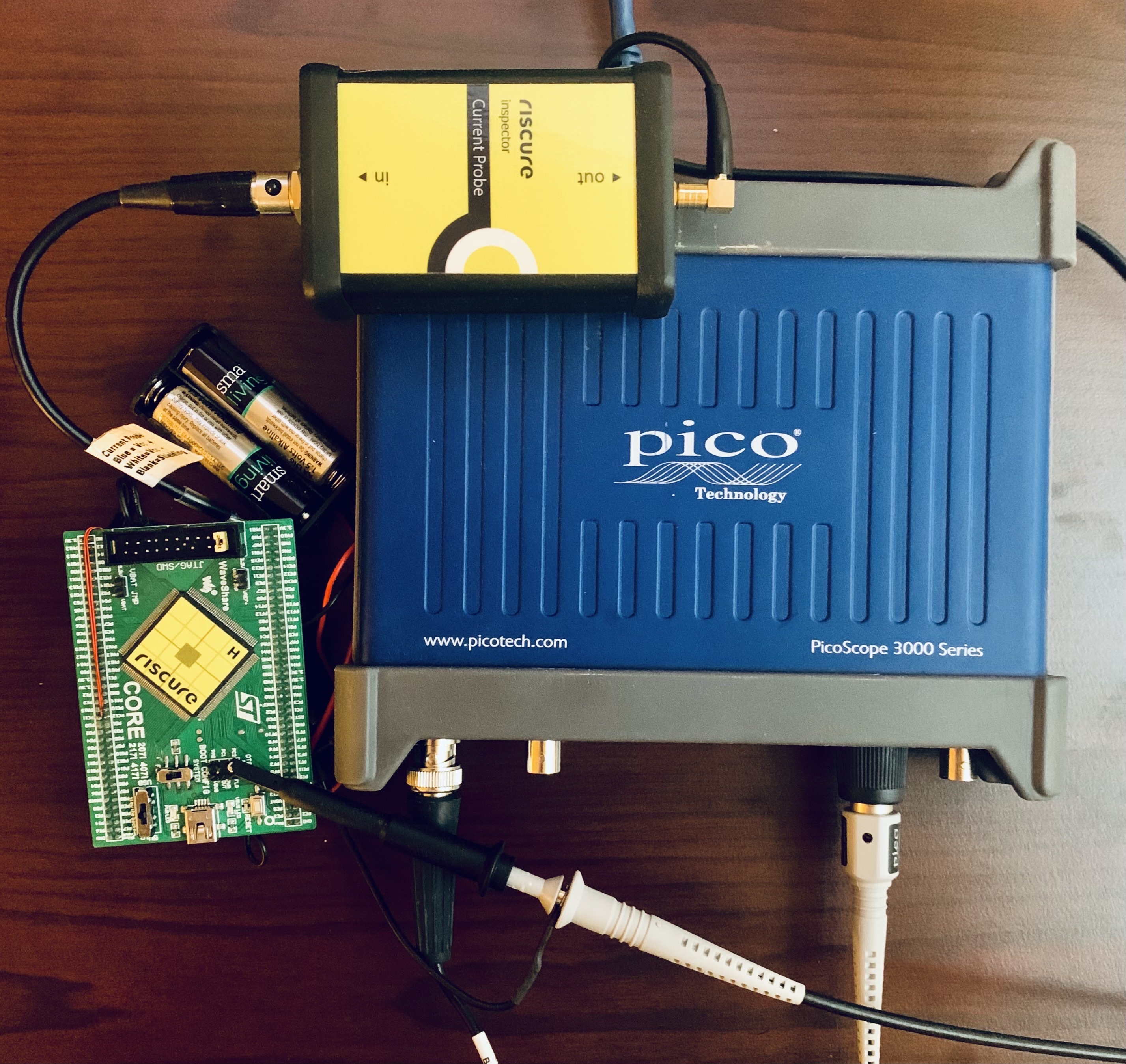}}
	\caption{Equipment for trace acquisition includes devices for measuring the power consumption of the target system. A current probe is used to capture the power consumption, outputting a voltage proportional to the measured power. The data is then transmitted to an oscilloscope, which digitizes the output for further analysis.}
    \vspace{1.5em}
	\label{fig:env}
\end{figure}

\section{Evaluating the Single-Trace Vulnerability}
\label{sec:res}
This section begins with a description of our measurement setup, followed by an analysis of the collected side-channel information and presentation of the results.
We begin by demonstrating that the attack distinguishes intermediate value assignments through graphical illustrations.
We then quantify our attack success rate using a theoretical model. 
Specifically, we employ a univariate Gaussian template~\cite{suresh_TA} with the selected point of interest (POI) to guarantee the success rate. 
Finally, we analyze the impact of the proposed attack on the security of FALCON.

\subsection{Measurement Setup}
\label{sec:Measurement_Stup}
\autoref{fig:env} shows our measurement setup. 
The target device is an ARM Cortex-M4F CPU operating at 30 MHz, which is a canonical setting for side-channel testing in embedded applications according to previous publications \cite{guerreau2022hidden,zhang2023improved,lin2025thorough, karabulut2021single, 
 guerreau2024not}. 
We selected the lowest supported frequency to reduce noise, as lower frequencies typically yield cleaner power traces.
Therefore, our measurements are not inherently limited by environmental noise.
We utilized the submission package from the NIST reference software implementation. 
Measurements were captured using a PicoScope 3206D oscilloscope set to a sampling rate of 250 MHz. 
A Tektronix CT1 passive current probe was used, offering a bandwidth of 1–1,000 MHz at 3 dB.
No external amplification was applied to enhance the measured signals. 

All experiments presented in this work were performed on an STM32F417 series board equipped with an ARM Cortex-M4 processor, a platform widely adopted in side-channel research due to its accessibility and relevance. 
Although extending the analysis to additional hardware targets would be a logical next step, we limit our evaluation to a single device to maintain experimental focus and reproducibility.
This decision is consistent with the methodology employed in several prior studies~\cite{kurian2025tpuxtract,lin2025thorough,karabulut2021single, guerreau2024not}.

\begin{figure}[t]
    \centering
    \includegraphics[scale=0.32]{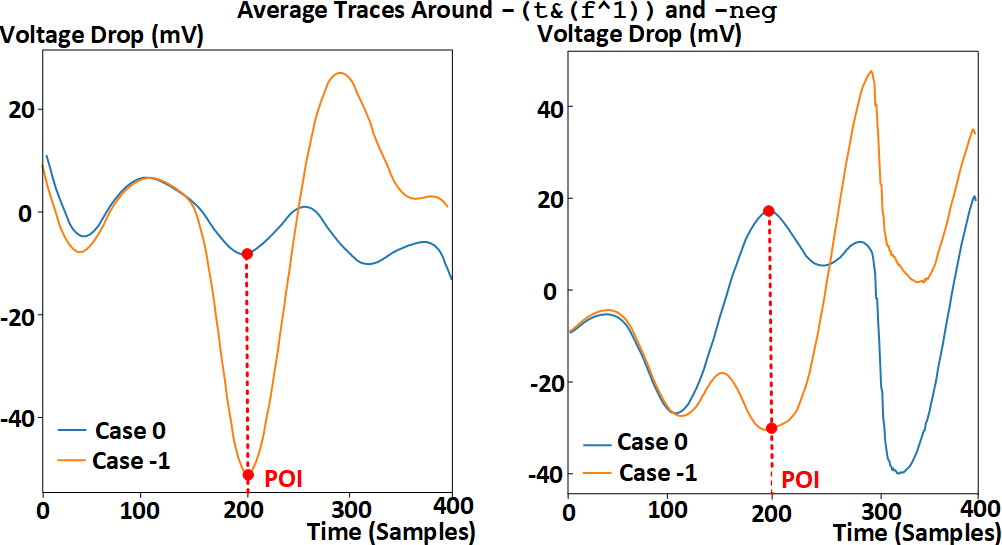}
    \caption{The average power trace around the leaky operations for the two attack points is depicted. The left figure corresponds to the first attack point, while the right figure represents the second attack point. At both attack points, only two possible cases exist: `0' and `-1'. The power consumption for the `-1’ case is observed to be higher than for the `0’ case.}
      \vspace{1.5em}
    \label{fig:Traces} 
\end{figure}

\subsection{Attack Results}
\label{sec:_attack}
\subsubsection{Results on \texttt{-(t \& (f \^{} 1))}}
The left panel in \autoref{fig:Traces} shows the average power trace of the 400 samples around the point-of-interest (POI) when attacking \texttt{-(t \& (f \^{} 1))}.
The horizontal axis represents time in samples, while the vertical axis indicates power consumption. 
Sample point 200 marks the POI identified in \autoref{sec:attack}. 
The results indicate that the average power consumption for the `-1' case is higher than for the `0' case.

We then quantify the attack success rate by modeling the power distribution to derive a theoretical estimate.
Since FALCON executes the targeted discrete Gaussian sampling step once to generate a single secret coefficient, we conduct a single-trace template attack. 

We selected the point of maximum correlation as the POI to build our univariate Gaussian template, though multiple POIs could be chosen to enhance the success rate on noisier platforms. 
At this POI, we computed the mean $\mu_i$ and variance of power \textit{$v_i$}.
\vspace{-.5em}
\begin{equation}
\label{eq:Pk}
    P_k = \sum_{j=0}^{k} \log \mathcal{N}(t_{j,{s_i}},\mu_i,v_i)\;\;,
\end{equation}

Using 500,000 measurements, we constructed the template by applying a normal probability density function (NPDF) at the POI. 
The computation incorporates the observed trace values $t_{j,{s_i}}$, along with the mean $\mu_i$, and variance $v_i$ obtained during profiling. 
To mitigate precision issues caused by extreme NPDF values, we computed the sum of log-likelihoods under the normal distribution $\mathcal{N}$, as shown in Equation \ref{eq:Pk}. 
The index of the matrix $P_k$ with the highest value corresponds to the predicted coefficient.

The left panel of \autoref{fig:Stat_Results} illustrates the Gaussian model derived from the data at the POI. The horizontal axis represents power consumption measured by voltage drop, while the vertical axis represents the probability density as derived from the template. The two bell-shaped curves correspond to the cases where the intermediate value is `-1' or `0'.
For this attack point, the two distributions are well separated, with overlapping area accounting for \( 2.56 \times 10^{-9} \%\) of the total area under the two curves, 
indicating a success rate greater than 99.999999999\%. The obtained Gaussian model was applied to 500,000 measurements to derive classification labels. A comparison with the ground truth shows a classification accuracy of 100\% on real-world data.
\vspace{+0.5em}

\subsubsection{Results on \texttt{-neg}}
We followed the same template-building approach to evaluate the attack results on \texttt{-neg}. 
The right panel of \autoref{fig:Traces} illustrates the separation between cases `0' and `-1' cases, as observed in the average power traces.
The horizontal axis represents time samples, while the vertical axis reflects power consumption. 
These average traces show a clear distinction between the two cases.

\begin{figure}[t]
	\centerline{\includegraphics[scale=0.32]{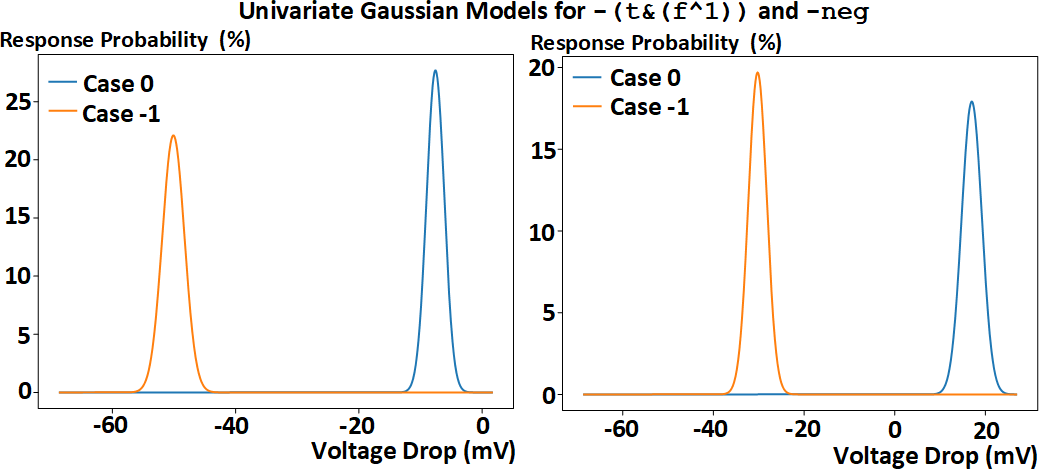}}
	\caption{The final results for the two attack points are presented. The left figure illustrates the results for the first attack point, while the right figure shows the results for the second attack point. We applied a univariate Gaussian model. The results are clearly separated in both cases, with no overlap, indicating clear and successful classification.}
	\label{fig:Stat_Results}
      \vspace{2em}
\end{figure}

The right panel of \autoref{fig:Stat_Results} illustrates the Gaussian model obtained from the first step of our attack. 
We selected the point of maximum correlation as the POI to build our univariate Gaussian template.
The horizontal axis represents power consumption, and the vertical axis denotes the response probability from the constructed template. 
The two bell-shaped curves illustrate the two cases, which are well separated at this attack point. The overlap accounts for only \( 2.55 \times 10^{-10} \%\) of the total area under the curves, indicating classification accuracy higher than 99.9999999999\%.
The Gaussian model was applied to 500,000 measurements, and the resulting classifications matched the ground truth, yielding an accuracy of 100\%.

\subsection{The Impact on FALCON’s Security}
Based on the generalized success rate for each secret value assignment described above, we analyze the impact on FALCON's security scheme. 
Each discrete Gaussian sampling subroutine execution involves running the outer loop twice and running the inner loop 52 times. 
Our attack on \texttt{-(t \& (f \^{} 1))} accomplished 99.999999999\% accuracy, while the attack on \texttt{-neg} accomplished 99.9999999999\% accuracy. 
Consequently, our attack for extracting one coefficient in FALCON is as follows: 

\[
\small \left( 99.999999999\% \right)^{52} \times \left( 99.9999999999\% \right)^{2}=99.9999999478\%
\]

Based on the success rates described above, the overall success rate for recovering the secret polynomials $f$ and $g$, in the FALCON-512 parameter set is as follows:
\[ 
\left((99.9999999478\% \right)^{512})^{2} = 99.99994654\%
\] 

For FALCON-1024, the overall success rate for recovering the secret polynomials $f$ and $g$ is as follows:

\[ 
\left((99.9999999478\% \right)^{1024})^{2} = 99.99989309\%
\]
We assert that this vulnerability represents a significant compromise of the FALCON cryptographic scheme.
\vspace{1em}
\section{Discussions}
\label{sec:disc}
In this section, we briefly discuss potential defense methods and calibration parameters considered in our experimental setup. 
We also analyze the limitations of the proposed attack.

\subsection{Defense Methods}
As in prior studies~\cite{kurian2025tpuxtract,karabulut2021falcon,karabulut2021single}, this work demonstrates a practical single-trace attack to highlight real-world risks associated with vulnerabilities in FALCON and to promote awareness within the developer community. 
We offer a brief overview of potential countermeasures without emphasizing any specific approach.
The design and implementation of these measures are considered outside the scope of this paper.

Defenses against single-trace side-channel vulnerabilities can be implemented at both the hardware and software levels. 
On the hardware side, constant-power designs aim to flatten the power profile of cryptographic operations, thereby preventing power traces from revealing data-dependent behavior. 
Such designs can be implemented through custom circuitry that ensures uniform power consumption.
For example, a switched-capacitor power supply can be employed as a hardware-level countermeasure~\cite{10752788}. 
This technique charges a bank of capacitors during non-sensitive operations and uses the stored energy to power the device during sensitive computations—such as the discrete Gaussian sampling subroutine—thereby reducing leakage. 
As a result, the device draws minimal or constant power from the external supply during these critical periods, reducing observable leakage.
While effective, these approaches often require significant changes to the hardware architecture and may incur performance and area overheads.

On the software side, a widely used class of countermeasures is known as hiding, which seeks to obscure the relationship between internal computations and side-channel leakage.
This can be achieved by inserting dummy operations, introducing random delays, or reordering independent instructions, thus reducing the temporal correlation between power consumption and specific intermediate values \cite{cryptoeprint:2022/1416}. 

\subsection{Applicability to Other Implementations}
The proposed attack also applies to other algorithms that negate the result of a 63-bit right shift applied to 64-bit intermediate variables containing secret data.
The identified vulnerability exists in all of FALCON’s software implementations. 
This includes both the baseline and optimized versions provided in the NIST submission package, as the discrete Gaussian sampling subroutines remain unchanged across all implementations.
Although the attack was demonstrated on FALCON-512, it applies equally to FALCON-1024, which employs the same sampling subroutine.

\subsection{Calibration Factors}
The platform's noise level decreases as the device's operating frequency is lowered. 
To minimize measurement noise, we configured the development board to operate at its minimum supported frequency of 30 MHz, in line with prior studies~\cite{aysu18horizontal, karabulut2021single, owens2024efficient}. 
Earlier works have demonstrated single-trace side-channel attacks at even lower frequencies, such as 8 MHz in attacks on the NTT~\cite{second_NTT_attack}.
Analyzing higher frequencies may require more sophisticated power measurement equipment, additional probes for near-field electromagnetic leakage detection, or amplification and post-processing for noise reduction.

Although full key recovery demonstrations can be informative, our evaluation focuses on real-world recovery of intermediate coefficients at two identified leakage points. 
We then use theoretical modeling to guarantee the overall success rate of full key extraction.
This follows a common practice in prior work~\cite{frodo_single_poly, karabulut2021single, first_template}, where full key recovery is omitted due to the repetitive nature of applying the template to power measurements and the limited analytical value it provides. 
In the case of FALCON, fully recovering the secret key requires applying the same obtained template to power measurements at least 104 times, providing limited insights.
More importantly, while strong performance on a specific test set may demonstrate success in an individual instance, only a theoretical model can provide generalizable guarantees for future measurement success and inference across varied conditions.

\subsection{Drawbacks of Our Attack}
Template attacks have well-known limitations and challenges, including processing time constraints. 
In our scenario, selecting a single POI resulted in a high success rate and required only a few minutes to construct the profiling templates.
On devices with more complex power profiles, incorporating additional POIs could further improve attack effectiveness, though it would also incur higher computational overhead.

Our paper follows the same method used in prior demonstrations of single-device attacks \cite{lin2025thorough, karabulut2021single, guerreau2024not}. 
For attacks to succeed on devices of different makes and models (also known as cross-device attacks), it is essential to develop device-specific power profiles that consider architectural features such as pipelining and out-of-order execution. 
Overcoming these challenges may require advanced machine-learning-based profiling techniques~\cite{kashyap20, abdelkhalek2021investigating} or reconstructing the profiling template on newly tested device types. 
It is important to recognize that the challenge of cross-device single-trace side-channel attacks on post-quantum cryptosystems remains an open problem. 
Our paper adopts the method used in prior demonstrations of single-device attacks~\cite{second_NTT_attack, frodo_single_poly, single_trace_18, ravi_dilit_sing}. 

\section{Conclusions}
\label{sec:conc}
While lattice-based cryptography offers strong post-quantum security with relatively low computational overhead, it relies on specialized operations that have not been extensively scrutinized for side-channel leakage. 
In this work, we revealed a critical, new vulnerability in software implementations of FALCON, specifically related to the negation of a value obtained via a 63-bit right shift. 

Our analysis demonstrated that FALCON’s discrete Gaussian sampling routine leaks intermediate value assignments, enabling full recovery of the secret key. 
The attack is validated on an off-the-shelf embedded device running both the reference and optimized implementations from FALCON's NIST submission package, confirming its practicality. 
Notably, the uncovered vulnerability differs from known single-trace attacks and is inherently distinct from multi-trace approaches.
Consequently, existing countermeasures designed for other types of leakage may be ineffective and needs to be re-evaluated in light of these findings and modified if needed. 
This paper, therefore, presents a concrete attack, with the primary goal of highlighting the specific risks of using FALCON's unprotected reference software in key generation and informing the developers of this leakage.


{\selectfont{\bibliographystyle{IEEEtran}}
\vspace{-.5em}

\vspace{-5em}
\end{document}